\documentclass[aps,twocolumn,nofootinbib,longbibliography]{revtex4-1}
\usepackage[babel]{csquotes}
\usepackage{graphicx}
\usepackage{amsmath,amssymb}
\usepackage[colorlinks,citecolor=blue,linkcolor=blue,urlcolor=blue]{hyperref}
\usepackage{mathrsfs}
\usepackage{enumerate}
\def\be{\begin{equation}}
\def\ee{\end{equation}}

\begin{document}

\title{An argument against the realistic interpretation of the wave function}

\author{Carlo Rovelli\vspace{.5em} }
\affiliation{\small
\mbox{CPT, Aix-Marseille Universit\'e, Universit\'e de Toulon, CNRS,} \\
Case 907, F-13288 Marseille, France. \\ 
and \\ 
 \mbox{Samy Maroun Research Center for Time, Space and the Quantum.\vspace{.3em}
}}

\date{\small\today}

\begin{abstract}
\noindent  
Testable predictions of quantum mechanics are invariant under time reversal. But the change of the quantum state in  time is not so, neither in the collapse nor in the no-collapse interpretations of the theory. This fact challenges the realistic interpretation of the quantum state. On the other hand, this fact follows easily if we interpret the quantum state as a mere calculation device, bookkeeping {\em past} real quantum events. The same conclusion follows from the analysis of the meaning of the wave function in the semiclassical regime.  

\end{abstract}

\maketitle

In classical mechanics, we cannot figure out the arrow of time from a given history.\footnote{Unless initial data are highly non generic --for instance have low entropy.}  This is a consequence of the invariance of classical mechanics under time reversal.

Quantum mechanics has the same property, in the following sense. If we observe the value $a$ of an observable\footnote{Throughout the article I assume for simplicity all eigenvalues to be non degenerate.} , the probability of observing the value $b$ of another observable after a time $t$  is equal to the probability of observing the value $a$  a time $t$ after $b$ was observed (because $|\langle a|e^{-iHt}|b\rangle|^2=|\langle b|e^{iHt}|a\rangle|^2$).\footnote{Penrose observes in \cite{Penrose2005} that $T$-invariance is broken by two detectors $d_1,d_2$ separated by a beam splitter, because $P_{1\to2}=\frac12$ while $P_{2\to1}=1$. But this is a consequence of the asymmetric assumption that in reverse time the beam necessarily gets to $d_1$: the state is fully specified at $d_2$ but not at the $d_1$.} More precisely: given an ensemble of sequences of measurements outcomes, we cannot figure out the arrow of time from them.\footnote{Unless the initial ensemble is highly non generic --for instance has low entropy.}   

However, this T-invariance is badly broken in the textbook accounts of quantum theory. To see this, consider  a system initially in a state $|a\rangle$, eigenstate of an operator $A$ with eigenvalue $a$.  Say at time $t_a$, we have measured the value $a$ for the quantity represented by the operator $A$. At a later time $t_b$, we measure the value $b$ for a quantity represented by the operator $B$.   The actual measurement outcomes are thus $a$ at $t_a$ and $b$ at $t_b$.  
\begin{eqnarray}
              t_a  &\longmapsto& t_b \nonumber\\
              a  &\longmapsto& b.  \label{q}
\end{eqnarray}
If we reverse the history of the events, we have the time reversed history 
\begin{eqnarray}
              t_b  &\longmapsto& t_a \nonumber\\
              b  &\longmapsto& a.  \label{qe}
\end{eqnarray}
Nothing remarkable so far.  But let us now time reverse not just the quantum events, but rather the full evolution of the state (or wave function). According to the standard account, at time $t_b$ the state of the system is projected to an eigenstate of $B$, say $|b\rangle$, with eigenvalue $b$.   Assuming for simplicity the system was already in $|a\rangle$ to start with, we have 
\begin{eqnarray}
              t_a  &\longmapsto& t_b \nonumber\\
              |a\rangle  \longmapsto& |a\rangle&  \longmapsto |b\rangle  . 
              \label{primo}
\end{eqnarray}
If we time reverse this we obtain
\begin{eqnarray}
              t_b  &\longmapsto& t_a \nonumber\\
              |b\rangle  \longmapsto& |a\rangle &  \longmapsto  |a\rangle   \label{trr}
\end{eqnarray}
which is \emph{not} what textbook quantum theory predicts! The evolution of the quantum state  predicted by standard quantum theory in association with the quantum events \eqref{qe} is rather 
\begin{eqnarray}
              t_b  &\longmapsto& t_a \nonumber\\
              |b\rangle  \longmapsto& |b\rangle&  \longmapsto |a\rangle,     \label{qe2}
\end{eqnarray}
which differs from  \eqref{trr} because the state of the system in between the two measurements is $ |b\rangle $, rather than  $|a\rangle$.  Therefore:  {\em The quantum events predicted by quantum theory (the outcomes of measurements) and their relative probabilities are T-invariant.  But the evolution of the quantum states (the wave function) in the conventional accounts of quantum mechanics is not.} 

A movie of a classical evolution projected backward is still a consistent classical evolution.  A movie with the sequence of the observed quantities in a series of quantum measurements is still a consistent sequence of observations. {\em But a movie of the changes of the wave function evolution projected backward makes no sense.}

This can be put pictorially: we can describe the $\beta$-decay of a nucleus by means of the electron wave function concentrated on the nucleus, slowly leaking out in all directions until the electron gets detected by a Geiger counter at some distance.  The time reversed phenomenon --the Geiger apparatus emits an electron that is then captured by the nucleus-- is perfectly possible, but it is not described by a wave function that converges symmetrically onto the nucleus. 

Notice that in all these cases there exists a consistent quantum  description of the time reversed sequence of events (which is \eqref{qe2}), but at a given time between two observed events the quantum state is different depending on which arrow of time we assume to be the ``actual" one. But if Nature knows nothing about the ``actual" arrow of time, how can the wave function know which one is the correct choice?

The breaking of T-invariance is due to the common assumption that the state ends up to be  projected \emph{after} a measurements, and not \emph{before}. While at first sight this assumption sounds very plausible (we say that effects come after their causes, not before), at a more careful examination this is in striking contrast with anything happening in classical mechanics, where all arrows of time can always be traced back to special (low entropy) initial conditions in the past. Here we appear to have a striking violation of T-invariance in the dynamics of the state, unrelated to any peculiar initial condition. 

To put it simply: between a measurement of $A$ and a measurements of $B$ what determines if the state is $|a\rangle$ or $|b\rangle$?  

What is truly striking, however, is that such a gross violation of T-invariance turns out to be unobservable! Because the actual sequence of measurements knows nothing about whether state between the measurements is $|a\rangle$ or $|b\rangle$.  The situation is a bit like in the official Catholic doctrine, according to which \emph{two} miracles happen in the Mass: bread is mutated in flesh, but flesh miraculously looks like bread.  Here, the quantum state knows about the direction of time, but miraculously it keeps this info for itself. 

This breaking of T-invariance at the core of the common textbook presentation of quantum mechanics has been emphasized by Huw Price in his remarkable book about the arrow of time \cite{Price}, where it is discussed in great detail. I refer the reader to the book for a discussion and replies to  possible objection to the above.  In the next section I discuss what to make of it. 

\section{The two interpretations of the quantum state}

There are two main interpretations of the quantum state $|\psi\rangle$ of a physical system. 
\begin{enumerate}[(i)]
\item The first is that  $|\psi\rangle$  represents the actual state of affairs of the world. This is the {\em realist} interpretation of  $|\psi\rangle$, which can be traced back to Schr\"odinger.  This point of view is common in most textbook presentations of quantum theory.  It is also shared by several no-collapse interpretations, on which I will comment in the next section. 
\item The second is that $|\psi\rangle$ is just a theoretical bookkeeping devise for something else. In particular, it can be a way of bookkeeping the real events we are aware of. This point of view can be traced back to Heisenberg and Dirac. Example of an interpretation based on it are Einstein's claim that quantum mechanics is incomplete, or (maintaining the idea that quantum theory is complete), the Relational interpretation \cite{Rovelli:1995fv,Rovelli:1998mi,Smerlak:2006gi,Fraassen:2010fk,Dorato2013a}, where the actual ontology of the world is reduced to the sequence of the (relational) elementary quantum events $a$, $b$ ... that  happen when physical systems interact. 
\end{enumerate}
In this second family of interpretations, the fact that the evolution of the state is not time reversal invariant generates no concerns:  if the state is a bookkeeping device of {\em past} event, the ambiguity in the value of the state between $t_a$ and $t_b$ reflects our choice of bookkeeping past, or future, events.\footnote{We ourselves function entropically, namely exploiting past low entropy, therefore it is not surprising that we break $T$  invariance. }  But if the state is taken to be real, the fact that it behaves in a non T-invariant way, when everything we measure about the (classical and quantum) world is T-invariant, sounds illogical.   

Therefore the breaking of $T$ invariance in the changes of the state is a strong indication against a realistic interpretation of this state.  

The fact that assigning a state to the system between two measurements forces us to choose a direction of time --when physics is blind to the direction of time--, strongly indicates that the state of a system at time $t$ is not something about the actual physics at $t$: it is about what we know at that time.

\section{No-collapse interpretations}

There are interpretations of quantum mechanics based on a realistic interpretation of $|\psi\rangle$ where the collapse is assumed not to happen.  At $t_b$, the wave function of the system is not projected to $|b\rangle$.   The two main interpretations of this group are the Many Worlds interpretation  (MW) \cite{Saunders2010} and the de Broglie-Bohm hidden variable theory (dBB) \cite{Goldstein2001}. These interpretations assume that during a standard measurement what happens is only the entanglement between the system and the apparatus. 
\begin{eqnarray}
              &t_b&  \\ \nonumber 
              |a\rangle\otimes |o_i\rangle  &\longmapsto&  |b\rangle\otimes |o_a\rangle +  |b'\rangle \otimes |o_{b'}\rangle.
\end{eqnarray}
In MW, the result of the measurement is assumed to be realized in each of the branches (``worlds"). In dBB, the ontology includes also classical-like system's variables, whose dynamics is guided by $|\psi\rangle$; these happen to follow one or the other of the branches of the entanglement.  The ``empty" branch" ($ |b'\rangle \otimes |o_{b'}\rangle$) becomes de facto irrelevant if there is sufficient decoherence.

How are the two sequences of events \eqref{primo} and \eqref{qe2} described in this framework? The first gives (simplifying the notation of tensor states)
\begin{eqnarray}
              t_a : && \longmapsto  |ao_ao'_i\rangle +  |a'o_{a'}o'_i\rangle    \\ \nonumber 
              t_b:     
              &&    \longmapsto 
          |bo_ao'_b\rangle    +  |b'o_{a}o'{}_{b'}\rangle +|bo_{a'}o'_b\rangle    + | b'o_{a'}o'_{b'}\rangle   
\end{eqnarray}
while the second gives 
\begin{eqnarray}
              t_b : &&
                          \longmapsto  |bo_io'_b\rangle +  |b'o_{i}o'_{b'}\rangle    \\ \nonumber 
              t_a:     
              &&    \longmapsto 
          |ao_ao'_b\rangle    +  |a'o_{a'}o'_{b}\rangle +|bo_{a}o'_{b'}\rangle    + | a'o_{a'}o'_{b'}\rangle   
\end{eqnarray}
Again, during the intermediate between $t_a$ and $t_b$ is different in the two cases: in one case it is maximally entangled with $o$, in the other case it is maximally entangled with $o'$.   Therefore we are in the same situation as in the collapse interpretation: that is, the assumed evolution of $|\psi\rangle$ depends on the time orientation even if the empirically observable physics does not. 

One way to put this observation pictorially is to look at the cover of the book on MW \cite{Saunders2010}. See Figure 1. The cover image is a tree, and is meant to suggest the branching of the universal wave function.   Where does this branching comes from, if fundamental physics is invariant under time reversal?\footnote{A possible reply to these objections is to consider branching perspectival. The state of the universe is not branched and makes no distinction between past and future.  It is our choice that isolates a substructure of the true universal wave function, and it is only this one that appears to display the branching and its time arrow. In other words, the true state of the system we observe in the laboratory is highly entangled in the past as well as the future, but we ignore the aspects of this state which are not relevant to our interest in predicting \emph{future} evolution.  The price to pay for taking this view is that the wave function we use in a concrete utilizations of the theory is not the real wave function anymore.  It is a theoretical booking devise about some limited information we happen to have about the actual state of the world, because of some past interactions. But if so, then the ``real" state of WM and dBB is reduced to a hypothetical metaphysical entity about which we know very little; the quantity we call ``state" when we use quantum theory, instead, turns out to be precisely a mere bookkeeping device for our time-oriented limited information!  Along this line, MW and dBB start sounding like Lao Tzu \cite{LaoTzu1993}: ``The Wave Function that can be told of, is not the True Wave Function"...}
 
 \begin{figure}
\centerline{\raisebox{-1.7cm}{\includegraphics[width=3cm]{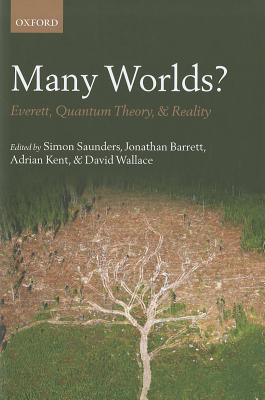}}\ \ \ or  \ \ 
{\raisebox{-1.7cm}{\includegraphics[width=3cm]{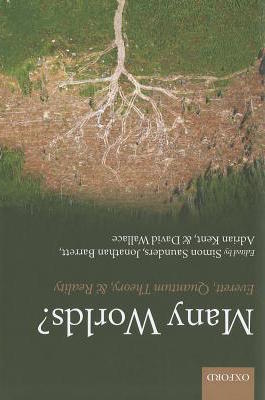}}}\ \ ?}
\caption{Where does the breaking of T-invariance comes from, in many worlds quantum theory?}
\end{figure}

\section{Price solution}

In  \cite{Price}, Huw Price chooses a realistic interpretation of the wave function, but explores a different solution to the problem of the breaking of T-invariance.  Price makes the hypothesis that there is a (hidden variable) state, well defined between any two measurements, which is not determined solely by the past measurement: it is determined also by the future measurement. This solution implies a form of retrocausality: what happens at time $t$ can be determined also by later events. 

This is different from classical theory, where what happens is entirely determined locally \emph{either} by the past \emph{or} by the future.  In Price proposal we have to assume that to determine locally what happens we have to know \emph{both} the past and the future.  Retrocausality is realized in classical mechanics as much as forward causality: since classical mechanics is T-invariant, the state of physical systems is equally determined by what happens before \emph{or} by what happens later. An arrow of time arises in classical mechanics only insofar uncommon \emph{past} conditions are realized.  In a relativistic theory, in particular, the state of the world on a spacial region $R$ determines uniquely everything in the future domain of dependence of $R$, \emph{as well} as everything in the past domain of dependence of $R$. Price suggestion, instead, is to build a hidden variable theory where the state between two measurements does not depend on one of the two, but rather on \emph{both}.   

The problem with this is that the state of the world at a given time, in a given region, is not sufficient to determine the state of the world in future, or past, domains of dependence.   This is a violent form of violation of locality, which, again, appears to contradict the empirical fact that, according to quantum theory, the probability distribution of all measurements performed in the domains of dependence of a spatial region $R$ \emph{are} uniquely determined by (sufficient) measurements in the region $R$.  Once again, one adds to the postulated real world a strange phenomenon, never observed in the universe, which however is then absent in the observed world!  Again like the real, but invisible, ``Flesh" of the note. 

If instead we renounce the idea that the quantum state represents the actual real state of the world at all times, and we interpret it simply as a way of coding what we happen to know about the (past or future) interactions of the system, then reality, much simplified, and formed simply by actual quantum events, is simply T-invariant, and subject to an entirely local dynamics.   This is for instance the ontology of the Relational interpretation.

\section{Conclusion}

The observed dynamics of the world is time-reversal invariant: a given un-oriented sequence of quantum events does not determine a time arrow.  Charging the wave function, (more in general, the quantum state) with a realistic ontological interpretation, leads to a picture of the world where this invariance is broken.  It is not broken in the sense that the full theory breaks $T$-reversal invariance (it does not), but in the sense that the wave function we associate to observed events depends on a choice of  orientation of time. 

Instead, restricting the ontology to the instantaneous (relational \cite{Rovelli:1995fv}) quantum events happening at interactions, saves us from such baroque constructions.  

In general, we can get clarity about the meaning of a quantity in an advanced theory by considering the role of this quantity in a regime where the theory reduces to physics we understand.  This can be done for the wave function by looking at the semiclassical limit of quantum mechanics.  In this limit, the wave function is the (exponent of the) Hamilton function. The Hamilton function is a quantity that nobody would dream of charging with an ontological meaning. It is a technical device for computing predictions about real events.  The same is true for the quantum state at some time. It is not a representation of reality at that time: it is a theoretical tool for computing probabilities about possible real quantum events.  

These quantum events provide a reasonable ontology for quantum theory.  Their characteristic aspect is 
to be discrete. Discreteness has been the first manifestation of quantum theory (quanta of light, discrete spectra, discrete angular momentum, particles in quantum fields...) and discreteness remains the core physical aspect of quantum theory: the number of distinguishable states in a finite phase space volume is finite --determined by the volume in $\hbar$ units. Forgetting discreteness and focusing only on the continuity of $\psi$, as a number of nowadays popular interpretations of quantum theory do, is misleading. Quantum mechanics, as the name  indicates, is about ``quantum" --that is ``discrete"-- events.  

God did not over-fill the world with moving waves on infinite dimensional configuration spaces. She didn't even draw it with heavy continuous lines.  She just dotted it with sparse quantum events.

 \begin{figure}[b]
\centerline{\includegraphics[width=5cm]{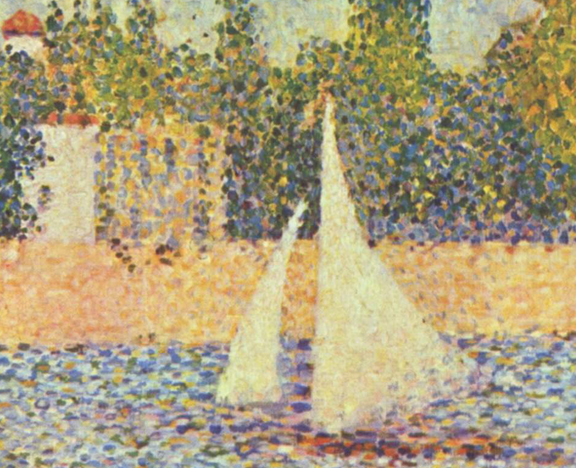}}
\caption{She just dotted it with sparse quantum events.}
\end{figure}

\providecommand{\href}[2]{#2}\begingroup\raggedright\endgroup

\end{document}